\documentstyle[12pt,aaspp4,flushrt]{article}

\def\etal{et al. }
\def\ch2{$\chi^2$ test}
\def\h{h^{-1}}
\def\bootes{Bo\"{o}tes }
\newcommand{\phmdeg}{\phm{\arcdeg}}
\newcommand{\phmmin}{\phm{\arcmin}}
\newcommand{\phmh}{\phm{$^h$}}
\newcommand{\phmm}{\phm{$^m$}}

\lefthead{El-Ad, Piran \& da Costa}
\righthead{AUTOMATED DETECTION OF VOIDS}

\slugcomment{ApJ Letters accepted}

\begin{document}

\title{Automated Detection of Voids in Redshift Surveys}

\author{Hagai El-Ad\altaffilmark{1}\authoremail{eladh@shemesh.fiz.huji.ac.il},
        Tsvi Piran\altaffilmark{1}\authoremail{tsvi@shemesh.fiz.huji.ac.il},
        and
        Luiz Nicolaci da Costa\altaffilmark{2,3}\authoremail{ldacosta@eso.org}}

\altaffiltext{1}{Racah Institute for Physics, The Hebrew University, Jerusalem, 91904 Israel}
\altaffiltext{2}{European Southern Observatory, Karl-Schwarzschild Stra{\ss}e 2, D-85748 Garching b. M{\"{u}}nchen, Germany}
\altaffiltext{3}{Observat\'orio Nacional, Rua Jos\'e Cristino 77, Rio de Janeiro, Brazil}

\begin{abstract}

We present a new void search algorithm for automated detection of 
voids in three-dimensional redshift surveys. 
Based on a model in which the main features of the LSS of the 
Universe are voids and walls, we classify the galaxies into wall
galaxies and field galaxies and we define voids as continuous 
volumes that are devoid of any wall galaxies. Field galaxies
are allowed within the voids.
The algorithm makes no assumptions regarding the shapes of the 
voids and the only constraint that is imposed is that the voids 
are always thicker than a preset limit, thus eliminating connections 
between adjacent voids through small breaches in the walls.  
By appropriate scaling of the parameters with the selection function 
this algorithm can be used to analyze flux-limited surveys. 
We test the algorithm on Voronoi tessellations and apply it to the 
SSRS2 redshift survey to derive the spectrum of void sizes and 
other void properties. We find that the average diameter of a void 
is $37\pm 8 \h$ Mpc.
We suggest the usage of this fully automated algorithm to derive a
quantitative description of the voids, as another tool in describing 
the large scale structure of the Universe and for comparison with
numerical simulations.

\end{abstract}


\clearpage
\section{Introduction}

Perhaps one of the most intriguing findings of dense and complete
nearby redshift surveys has been the discovery of large voids on 
scales of $\sim 50 \h$ Mpc and that such large voids appear to be
a common feature of the galaxy distribution (\cite{gel89,dc94}).
Deeper surveys support these findings (\cite{lan96}) with some even 
suggesting the presence of larger voids (\cite{bro90}). 
Since the discovery of the large \bootes void it has been recognized 
that inhomogeneities on such scales could impose strong constraints 
on theoretical models for the formation of large scale structure. 
However, if galaxy-voids reflect true voids in the underlying mass 
distribution, then despite the potential impact that this could have 
in constraining the power spectrum of mass perturbations (e.g., 
\cite{blu92,pir93}) or in our understanding of the biasing processes 
in galaxy formation, the voids have been largely ignored in recent 
work.

The major obstacle here has been the difficulty of developing proper
tools to identify voids in an objective manner and to quantify their
properties. We present a new void search technique for detection of
voids and the determination of their properties in an automated way,
not biased by the human eye. We test this algorithm using a Voronoi
tessellation and we apply it to the SSRS2 survey.

\section{The Void Searching Algorithm}

Our algorithm is based on a model in which the main features of the
LSS of the Universe are voids and walls. 
{\it Walls} are generally thin, 2D structures characterized by a high 
density of galaxies. Although coherent over large scales the walls are 
not homogeneous, and like our eyes we would like to ignore small 
breaches in the walls when identifying the individual voids. 
Galaxies within walls are hereafter labeled {\it wall galaxies}.  
The walls constitute boundaries between under-dense regions, generally
ellipsoidal in shape. These under-dense regions -- the so called
{\it voids} -- are not totally empty: there are a few galaxies scattered
here and there, and we will name these {\it field galaxies}.
 
Previous algorithms have used various definitions for voids: 
\cite{pel89} examined ensembles of contiguous cells with densities below 
a given threshold.{} 
\cite{kau91} used empty cubes, to which adjacent faces could be attached.{}
\cite{lin95} looked for single spheres, that are devoid of a 
certain type of galaxies. The latter two definitions constrain the shapes 
of the voids to be spherical or roughly ellipsoidal.  
We define a void as a continuous volume that does not contain any wall 
galaxies and is nowhere thinner than a given diameter. In other words, 
one can freely move a sphere with the minimal diameter all through the 
void. This definition does not determine the shape of the void: it can 
be a single sphere, an ellipsoid, or have a more complex shape, including
a concave one.

An important feature in our definition is that voids may contain field 
galaxies. A stiffer requirement, such that voids would be completely 
empty volumes, is too restrictive for it would imply that a single field
galaxy located in the middle of what we would like to recognize as a 
void might prevent its identification. However, for this definition to 
be practical we must be able to identify the field galaxies before we 
can start locating the voids.

Our algorithm is divided into two steps. First the {\it Wall Builder}
identifies the wall galaxies and the field galaxies. Then the {\it
Void Finder} finds the voids in the wall galaxy distribution. All
together, our method incorporates three parameters, which will be
defined below. Two of these parameters ($n$ and $\beta$) are used to
locate the field galaxies; the third parameter ($\xi$) is used during
the void search. Specific values for these parameters were chosen
after trials-and-errors with various simulations, to give results that
resemble as much as possible eye-estimates of the voids.
A detailed parametric study of the dependence of the algorithm on the
values of these parameters will be published elsewhere.

A wall galaxy is required to have at least $n$ other wall galaxies  
within a sphere of radius $L$ around it. Every galaxy that does not 
satisfy this condition is classified as a field galaxy. Note that the 
above requirement must be applied in a recursive manner until all the 
field galaxies are found.
Let the distance to the $n$'th closest neighbor of a galaxy be $l_n$.
For the sample this quantity has an average value $\bar{l}_n$ and a 
standard deviation $\sigma_n$. The radius $L$ is defined as 
\mbox{$L \equiv \bar{l}_n + \beta \sigma_n$}.
We have chosen here $n = 3$ and $\beta = 1.5$. Note that $n = 3$ is the 
minimal value that enables filtering of long thin chains of galaxies, 
generally preferring 2D structures that we would like to recognize as 
walls. 
As a side-bonus of this procedure we identify the walls, by connecting 
together all of the wall galaxies. These connections are not used in the 
next step but provide us with another visual tool to examine our results
(see Fig. 1, left panel). 
Hence this part of the algorithm is called the {\it Wall Builder}.
Note that smoothing is not required here since the statistics used are 
based on a galaxy count.

The {\it Void Finder} searches for spheres that are devoid of 
any wall galaxies. These spheres are used as building blocks for the 
voids. A single void is composed of as many (or as few) superimposing 
spheres as required for covering all of its volume. 
The limits of the distribution are treated as rigid boundaries, thus
causing some distortion in the voids found close to the boundaries: 
these voids tend to be smaller than the rest.
For a void with a maximal sphere of a diameter $d_{\max}$ we consider only
spheres with diameters larger than $\xi d_{\max}$, where $\xi$ is the  
``thinness parameter''. If the void is composed of more than one sphere 
(as is usually the case), then each sphere must intersect at least 
one another sphere with a circle wider than the minimal diameter
$\xi d_{\max}$. We have taken $\xi = 8/9$, which allows for enough 
flexibility without connecting distinct voids. A lower $\xi$ reduces 
the total number of the voids, with a slow increase in their total volume.

A major problem encountered is that of keeping apart neighboring voids.  
The walls separating voids often lack a few bricks, and thus a gap is 
created through which finger-shaped voids might find their way into a 
neighboring void, in this way connecting two that should have been kept 
separate. We use the following iterative procedure to overcome this 
problem. First, we look for the largest voids, using a large maximal 
sphere. Thus, voids found during the $i$'th iteration are composed of 
spheres in the diameter range \mbox{$\xi d_i < d < d_i$}, where $d_i$ 
denotes the maximal diameter $d_{\max}$ required in the $i$'th iteration
(voids containing spheres with a larger diameter having been detected 
during former iterations). Following iterations search for
smaller voids, avoiding regions in which voids were found by former
iterations. Once a void is found in a certain iteration, it is not
tampered with afterwards. Thus, voids found in later iterations can
not send ``thin fingers'' to previously identified voids.

The iterative nature of this method naturally brings up the question
of when to stop: we could go on until we mark every empty space as a
part of a void. The criterion we use here is based on comparisons
with Poisson distributions with the same geometry and the same
number of galaxies. Averaging together many such random simulations
one can derive $N_d$, the expectation value $N$ for the number of voids
found containing at least one sphere with a diameter larger than $d$.
As long as the actual number $\hat{N}_d$ of voids found in the original 
distribution exceeds $N_d$ (\mbox{$\hat{N}_d \gg N_d$}), the voids are 
considered significant and we proceed to the next iteration. 

The parameter determining the iteration in which a void is detected is 
the radius of the largest sphere contained in that void and not its 
total volume. Of course the total volume of a void and the radius of
the largest sphere contained in it are correlated, since voids with a 
bigger largest-sphere tend to be bigger voids. However, this may not 
always be the case: a void composed of a single large sphere will be 
detected earlier -- and hence considered more significant -- than a 
{\it larger} void composed of several spheres all having smaller radii.
This is in agreement with the fact that clear spherical voids are more 
prominent when a galaxy distribution is inspected by eye. In retrospect
this choice is also justified by the theoretical expectation that voids 
become more spherical with cosmological time (\cite{blu92}).

\section {Applications to Mock Surveys}

We have used Voronoi tessellations (\cite{vor08}) as a test bed for our 
algorithm. By construction Voronoi tessellations have the desired 
characteristics consisting of large empty regions and wall galaxies. 
To those we randomly add galaxies and an additional degree of 
stochasticity arises from the realization. The location and number of 
the Voronoi cells (the would-be voids), the spread of the wall galaxies 
and the amount of random galaxies are all known in advance. Therefore, we 
can test how well our algorithm performs in recovering the voids in the
tessellation.
 
Figure 1 shows a single slice cut through a cubic Voronoi tessellation. 
This tessellation was created using 8 nuclei for the Voronoi cells, with 
3000 galaxies of which 10\% were located randomly all over the volume. 
The designated wall galaxies were positioned randomly on the boundaries 
between the Voronoi cells, with a Gaussian displacement in the distance
from the cell boundary. 
The left panel depicts the original Voronoi cells, over which we superpose 
the simulated galaxy distribution. Note the walls highlighted along the 
boundaries between the Voronoi cells.
The right panel shows the voids, as identified by the Void Finder. 
Note how the field galaxies were filtered, thus allowing for a good fit 
between the Voronoi cells and the voids found.

\section{Voids in the SSRS2 Survey}

We have applied our algorithm to the recently completed SSRS2 survey 
(\cite{dc94}) which consists of $\sim 3600$ galaxies in the region 
\mbox{$-40 \arcdeg < \delta < -2.5 \arcdeg$} and $b \le -40 \arcdeg$.
In order to apply the algorithm to actual redshift surveys, one must
take into account the geometry and selection effects for a 
magnitude-limited sample.
Our algorithm can process any survey-like geometry defined in spherical
coordinates, treating the limits of the survey as rigid boundaries. 

In a magnitude-limited redshift survey like the SSRS2 the average
galaxy number density decreases with depth since only the more luminous
galaxies are visible at larger distances. If not corrected, this
selection effect will interfere with the algorithm: field galaxies will
occur more frequently, and the derived size of the voids will be larger 
because of the decrease in the mean density. All together, systematically 
larger voids will be found at larger distances. 

To minimize these effects we have considered a semi-volume limited sample 
consisting of galaxies brighter than $M_o \leq -19$, corresponding to a 
depth $r_o = 79.5 \h$ Mpc. We have computed the selection function:
\begin{equation}
\phi(r) = \frac { \Gamma (x_M, 1-\alpha) }
                { \Gamma (x_{M_o}, 1-\alpha) } 
\end{equation}
where $x_M = 10^{0.4(M_* - M)}$, using a Schechter luminosity function 
(\cite{sch76}) 
with \mbox{$M_* = -19.6$} and $\alpha = 1.2$ as derived for the SSRS2 
(\cite{dc95}).

We apply corrections based on the selection function to both phases of
the algorithm. In the Wall Builder phase, we consider larger spheres when
counting the neighbors of more distant galaxies. The volumes of the
counting-spheres are:
\begin{equation}
V = \left\{ \begin{array}{ll}
                  V_0              &        r < r_o \\
                  {V_0 / \phi (r)} &  r_o < r < r_{\max}
            \end{array}
    \right.
\end{equation}
where $V_0 = 4 \pi {L}^3 / 3$. The radius $L$ is calculated here using
only galaxies with $r < r_o$. Our final semi-volume limited sample consists 
of 1898 galaxies, extending out to \mbox{$r_{\max}=130 \h$ Mpc} where 
$\phi$ has dropped to 17\%.

A similar correction is applied to the Void Finder. Voids of a certain
size found in a low density environment are less significant than voids of 
the same size found in a high density environment. In order that all the 
voids found by the Void Finder in a given iteration will be equally 
significant we adjust the algorithm so that at a given iteration only 
relatively larger voids are accepted, if located at $r > r_o$. Thus we 
determine an initial diameter $d_{io}$ for the void search within the 
volume-limited region, and scale it by the selection function at larger 
redshifts in the same way as we change the volumes of the counting spheres. 
During consecutive iterations this initial value is reduced both in the 
volume-limited region as well as at large redshifts, again properly scaled 
by the selection function.

We identified twelve voids within the volume probed by the SSRS2. 
These are the most significant voids found, all of which having been 
detected during initial iterations for which voids were rarely found in 
equivalent random distributions. 
Five additional voids are still quite large, having volumes that exceed those
of a sphere with a diameter of $25 \h$ Mpc, but these were found during 
later iterations and therefor are less significant and were not 
considered in the calculations below.
The location and the characteristics of the identified voids are given in
\mbox{Table 1}. The top part of the table includes the 12 voids found
in the survey; the bottom part includes the 5 additional voids.
Column (1) lists the values of $d_{io}$ used while detecting each void.
The diameters given in column (2) are of a sphere with the same volume 
as the whole void, as is listed in column (3).
The center of the void given in columns (4)-(6) is defined as its
center-of-(no)-mass. The density contrast estimate listed in column
(7) was corrected for the average galaxy density at the same distance
as the center of the void. Finally, in column (8) we give the fraction 
of the total volume of the void covered by the single largest sphere 
contained in it. This value is typically $\sim 50\%$ of the total volume 
of the void. 

The average size of the voids in the SSRS2 survey as estimated from the 
equivalent diameters is \mbox{$\bar{d} = 37 \pm 8 \h$ Mpc}, lower but 
consistent with eye estimates of Geller \& Huchra (1989) and da Costa 
\etal (1994), which set this figure at $50 \h$ Mpc, and similar to the 
value of $38 \h$ Mpc obtained from the first zero-crossing of the 
correlation function (\cite{gol95}). 
The largest void found in the SSRS2 survey has an equivalent diameter
$d = 56.4 \h$ Mpc, making it comparable in volume to the large void
found in the \bootes (\cite{kir81}). The shape of this void
approaches that of an ellipsoid whose major axis is perpendicular to
the line of sight, located at: 
\mbox{$85 \h$ Mpc $ < r < 130 \h$ Mpc};
\mbox{$-25.0 \arcdeg < \delta < -2.5 \arcdeg$}; 
\mbox{$21^h < \alpha < 23.5^h$}. 
This void might actually be larger, since it is bounded by the limits of 
the SSRS2 survey. The average under-density within the voids was found to 
be \mbox{$\delta \rho / \rho \approx -0.9$}, a quite remarkable result 
showing how empty voids are of bright galaxies.

Figure 2 depicts a 3D representation of the voids in the SSRS2 survey.
The figure is striking and it unmistakably shows the propriety of
using the picture of a void-filled Universe to describe the observed
galaxy distribution.  The twelve voids comprise 41\% of the survey's
volume. An additional 9\% is covered by the five less significant
voids, and more volume is covered by still smaller voids. We estimate
that the walls occupy less than 25\% of the volume of the Universe. It
is important to point out that because of the paucity of large clusters 
and the small amplitude of peculiar motions in the volume surveyed by 
the SSRS2, redshift distortions are small (\cite{dc95}) and the 
properties derived here should reflect those of voids in real space.

Figure 3 shows a histogram of two distributions of equivalent-diameter
spheres: the distribution of the 12 voids found in the SSRS2 survey, 
compared to several random distributions, averaged together. Most of 
the voids found in the random distributions have diameters smaller 
than $25 \h$ Mpc. We used a \ch2 to compare the distribution of the 
SSRS2 survey voids' diameters with that of the random distributions.  
The probability that the two originate from the same distribution is 
$P = 0.087$. Note that in any case the probability could not be smaller 
than $P = 0.035$ (the result of the \ch2 comparing these 12 voids to a 
null histogram, with the same number of degrees of freedom). We expect 
that when larger surveys are made available then (with more voids) the 
\ch2 will yield $P < 0.01$. For example, a survey with a volume twice 
as large (i.e., with 24 voids) would yield $P = 0.001$.

The method we applied, of volume-limiting a part of the survey, prevents 
us from directly benefiting from the excess information hidden with the 
faint galaxies that we have eliminated. However, we can use these 
galaxies to re-examine our results. There are 1264 such galaxies in the 
SSRS2 survey; obviously, they are all located at $r < r_o$. Almost 50\% 
of this region is covered by voids -- but only 10\% of the faint galaxies
are found within them. This verification is important, since the 
volume-limiting process chooses only the brighter galaxies. It is
interesting to compare the percentage of faint galaxies within voids, to
that of the brighter galaxies \mbox{$M_o \leq -19$} found within the
voids: only 5\% of the later kind are contained in the voids. This
result indicates that the population of faint galaxies within voids
is larger than the population of bright galaxies.
 
\section {Summary}

We have presented a new algorithm for the identification of voids
in redshift surveys. The algorithm mimics the human eye in identifying 
the voids, focusing on the voids and disregarding small breaches on the
walls. This feature, together with the flexibility in void shapes, 
the recognition and filtering of the field galaxies and the ability to 
correct for selection effects, makes the algorithm powerful and of 
interest for cosmological applications.

We have applied the algorithm to the SSRS2 data sample and found 12
voids with $\bar{d} = 37 \pm 8 \h$ Mpc, in good agreement with visual 
interpretation of these maps.
However, the algorithm is automated and objective and it 
identifies voids and computes their properties in a quantitative way.
The formalism  can be  applied also to mock 
surveys generated from N-body simulations. Therefore it should be 
considered a new tool in our arsenal for investigating the nature of 
clustering in the Universe. We hope that the results obtained in this 
way will prove useful in future comparisons between theory and data, 
and would allow the most remarkable feature of the observed galaxy 
distribution to be used in a quantitative way to constrain theories of 
large scale structure.

\acknowledgments
We would like to thank Shai Ayal for helpful discussions and comments.
We would also like to thank Rien Van de Weygaert for providing us with 
his Voronoi tessellation code. 
One of us (LNdC) would like to thank the Hebrew University for hospitality 
while part of this research was done.

\clearpage

\begin{deluxetable}{cccrrrcc}
\tablecolumns{8}
\tablewidth{0pc}
\tablecaption{Properties of the voids in the SSRS2 survey}
\tablehead{
\colhead{Largest} & \colhead{Equivalent} & \colhead{Total} & \multicolumn{3}{c}{Location of Center} & \colhead{Void} & \colhead{Largest} \\
\cline{4-6} 
\colhead{Diameter\tablenotemark{a}} & \colhead{Diameter} & \colhead{Volume} & \colhead{$r$} & \colhead{$r.a.$} & \colhead{$dec.$} & \colhead{Under-} & \colhead{Sphere's} \\
\colhead{[$\h$Mpc]} & \colhead{[$\h$Mpc]} & \colhead{[$h^{-3}$KMpc$^3$]} & \colhead{[$\h$Mpc]} & \colhead{[hours]} & \colhead{[degrees]} & \colhead{density} &
\colhead{Fraction} \\
\colhead{(1)} & \colhead{(2)} & \colhead{(3)} & \colhead{(4)} & \colhead{(5)} & \colhead{(6)} & \colhead{(7)} & \colhead{(8)} }
\startdata
35.1 & 42.1 & 39.3 &  76.5 & 1$^h$50$^m$ & -16\arcdeg47\arcmin & -0.89 & 0.54 \nl
29.9 & 43.4 & 42.8 &  90.8 & 3\phmh43\phmm & -30\phmdeg04\phmmin & -0.89 & 0.44 \nl
28.9 & 56.4 & 94.0 & 108.7 & 22\phmh 21\phmm & -13\phmdeg 08\phmmin & -0.91 & 0.31 \nl
27.6 & 33.3 & 19.2 &  70.2 & 21\phmh 40\phmm & -13\phmdeg 56\phmmin & -0.90 & 0.50 \nl
26.0 & 32.2 & 17.7 &  53.0 & 23\phmh 48\phmm & -24\phmdeg 39\phmmin & -0.94 & 0.49 \nl
26.0 & 30.4 & 14.9 &  56.1 &  3\phmh 48\phmm & -20\phmdeg 19\phmmin & -0.91 & 0.55 \nl
23.7 & 25.2 & \phn8.3 & 77.2 & 3\phmh 17\phmm & -11\phmdeg 40\phmmin & -0.91 & 0.73 \nl
22.6 & 27.8 & 11.3 &  85.3 & 23\phmh 17\phmm & -12\phmdeg 19\phmmin & -0.94 & 0.59 \nl
21.6 & 39.8 & 33.4 & 115.8 &  3\phmh 08\phmm & -14\phmdeg 18\phmmin & -0.92 & 0.50 \nl
20.8 & 38.2 & 29.3 & 102.7 &  0\phmh 35\phmm &  -9\phmdeg 28\phmmin & -0.73 & 0.30 \nl
20.8 & 29.4 & 13.2 &  86.2 &  0\phmh 44\phmm & -28\phmdeg 52\phmmin & -0.98 & 0.35 \nl
19.8 & 42.6 & 40.5 & 115.1 &  0\phmh 24\phmm & -29\phmdeg 03\phmmin & -0.89 & 0.31 \nl
\hline
19.2 & 33.0 & 19.0 & 114.5 & 2$^h$ 02$^m$ & -9\arcdeg09\arcmin & -0.73 & 0.58 \nl
19.2 & 26.8 & 10.0 &  73.4 & 22\phmh 57\phmm & -32\phmdeg 07\phmmin & -0.97 & 0.34 \nl
19.2 & 31.7 & 16.9 & 114.4 &  2\phmh 42\phmm & -33\phmdeg 03\phmmin & -0.75 & 0.71 \nl
18.5 & 33.8 & 20.4 & 112.2 &  4\phmh 17\phmm & -15\phmdeg 27\phmmin & -0.96 & 0.43 \nl
16.9 & 27.6 & 11.0 & 116.1 & 21\phmh 24\phmm & -33\phmdeg 17\phmmin & -0.75 & 0.84 \nl
\tablenotetext{a}{Identical values in several rows imply that these voids were all identified during the same iteration.}
\enddata
\end{deluxetable}

\clearpage

\clearpage
\plotone{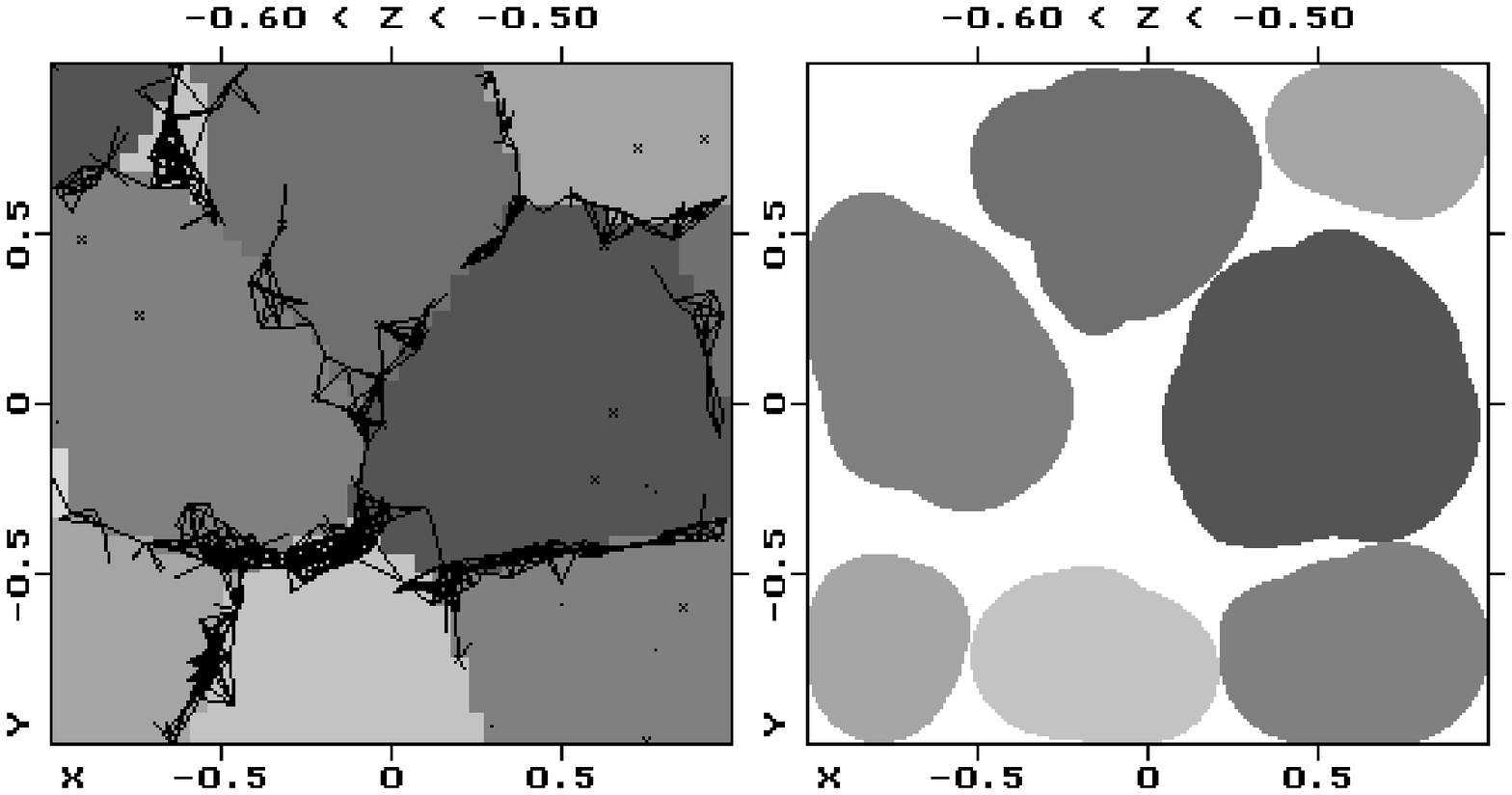}
\figcaption[fig1.ps]{A slice through a Voronoi tessellation. The width of the 
            slice shown here is 5\% of the whole cube. Left panel: the original
            Voronoi cells; each cell is drawn with a different color, in the 
            background. On top of the cells is the simulated galaxy 
            distribution. The walls are highlighted by connecting (black lines)
            the galaxies that the Wall Builder has identified as wall galaxies.
            The remaining, unconnected galaxies, are the field galaxies.
            `$+$' marks the galaxies distributed around the Voronoi boundaries.
            `$\times$' marks galaxies distributed randomly all over the volume.
            Right panel: the voids as found by our algorithm. Here each shaded
            area is a cut through a void at the middle of the slice.
            See text for more details.}

\clearpage
\plotone{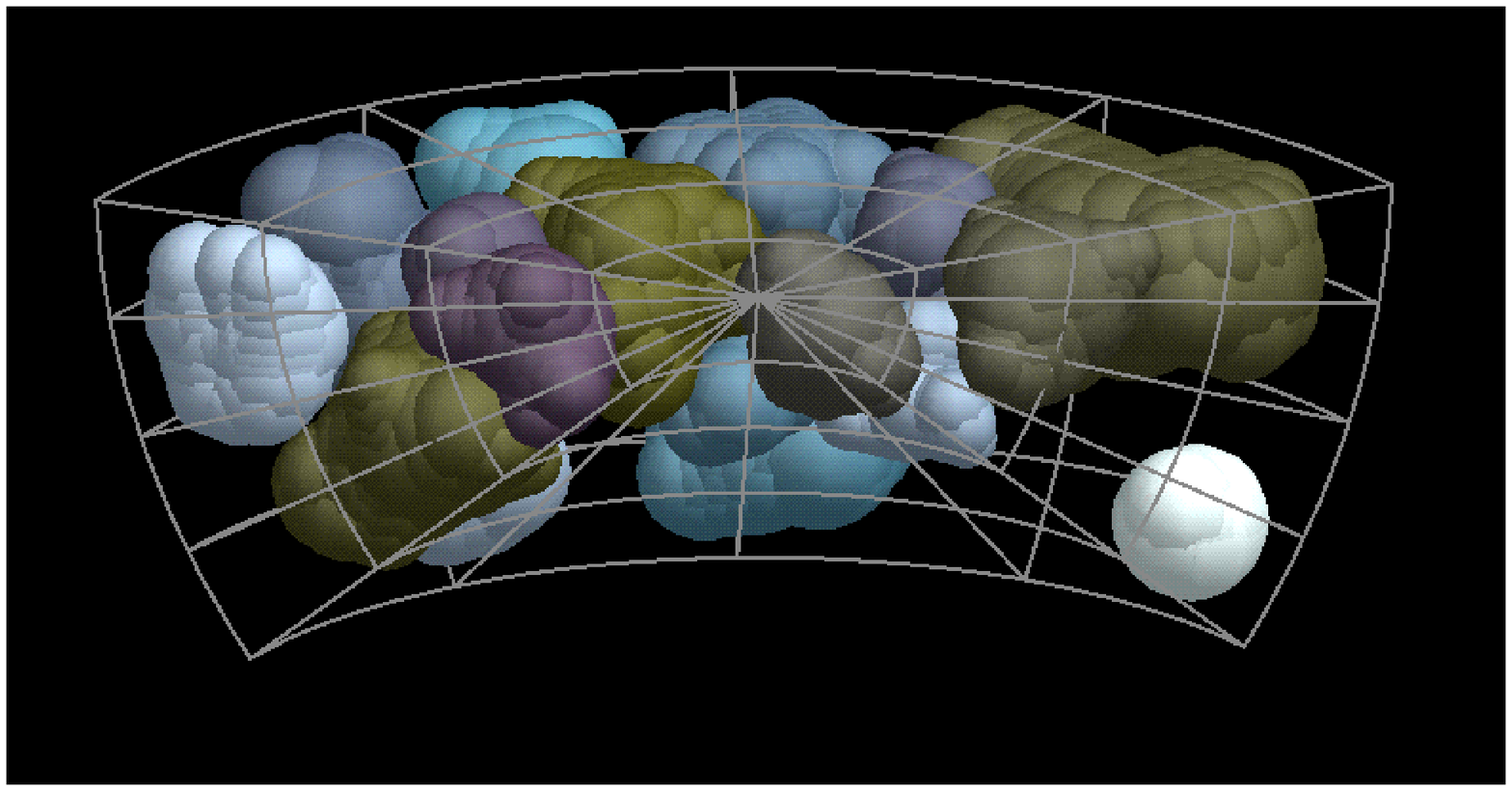}
\figcaption[fig2.ps]{3D view of the voids in the SSRS2 survey.}

\clearpage
\plotone{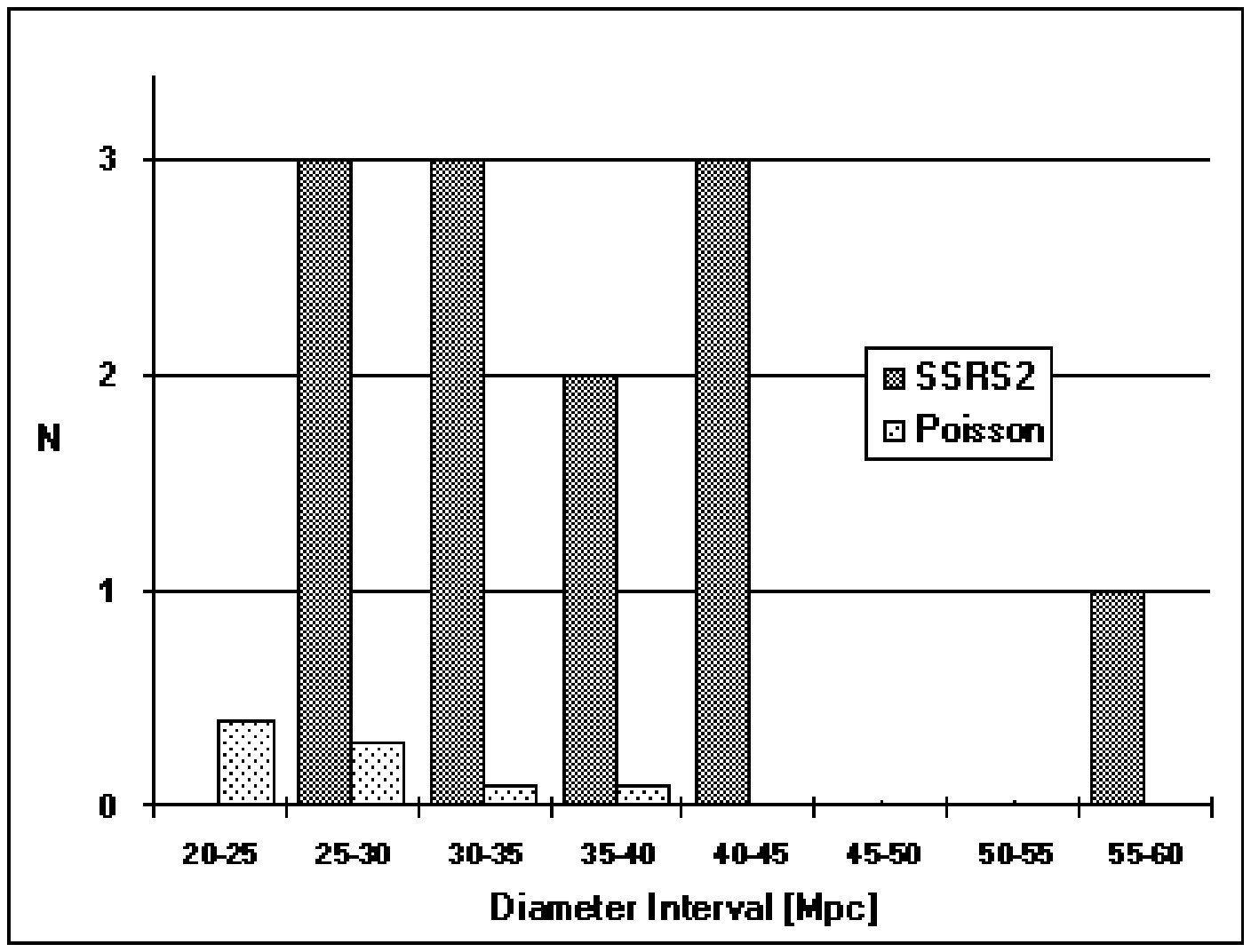}
\figcaption[fig3.ps]{Spectrum of void sizes in the SSRS2 survey, compared to 
            several equivalent random distributions, averaged together. The
            equivalent distributions have the same number of galaxies, the 
            same geometry and the same selection function as the SSRS2.}

\end{document}